\documentclass{aa}  

\usepackage{graphicx}
\usepackage{color,xcolor}
\usepackage[varg]{txfonts}
\begin{document}
   \title{Variable stars in the globular cluster NGC 7492.\thanks{Based on observations collected at the Indian Astrophysical Observatory,
Hanle, India.}}

   \subtitle{New discoveries and physical parameters determination.}
   
   \author{R. Figuera Jaimes
           \inst{1,2}
           A. Arellano Ferro
           \inst{3}
           D.M. Bramich
           \inst{1}
           Sunetra Giridhar
           \inst{4}
           \and
           K.Kuppuswamy
           \inst{4}}
           
   \titlerunning{Variables in NGC 7492.}

   \institute{European Southern Observatory, Karl-Schwarzschild-Stra$\beta$e 2, 85748 Garching bei M\"{u}nchen, Germany.\\
               \email{rfiguera@eso.org, robertofiguera@gmail.com, dan.bramich@hotmail.co.uk}
        \and
              SUPA, School of Physics and Astronomy, University of St. Andrews, North Haugh, St Andrews, KY16 9SS, United Kingdom.\\            
        \and
             Instituto de Astronom\'ia, Universidad Nacional Aut\'onoma de M\'exico, Apdo. Postal 70-264, M\'exico D. F. CP 04510, M\'exico.\\
              \email{armando@astro.unam.mx}
        \and
             Indian Institute of Astrophysics, Bangalore, India.\\
             }

   \date{Received November 5, 2012; accepted December 1, 2012}

 
  \abstract
   {}
   {We have performed a photometric $V,R,I$ CCD time-series analysis with a baseline of $\sim$8 years of the outer-halo globular cluster NGC 7492
with the aim of searching for new variables and using these (and the previously known variables) to determine the physical parameters of interest for
the cluster (e.g. metallicity, absolute magnitude of the horizontal branch, distance, etc.).}
   {We use difference image analysis (DIA) to extract precise light curves in the relatively crowded star field, especially towards the densely
populated central region. Several approaches are used for variability detection that recover the known variables and lead to new discoveries. We
determine the physical parameters of the only RR0 star using light curve Fourier decomposition analysis.}
   {We find one new long period variable and two SX Phe stars in the blue straggler region. We also present one candidate SX Phe star which requires
follow-up observations. Assuming that the SX Phe stars are cluster members and using the period-luminosity relation for these stars, we estimate
their distances as $\sim$25.2$\pm$1.8 and 26.8$\pm$1.8~kpc, and identify their possible modes of oscillation. We refine the periods of the two
RR Lyrae stars in our field of view. We find that the RR1 star V2 is undergoing a period change and possibly exhibits the Blazhko effect. Fourier
decomposition of the light curve of the RR0 star V1 allows us to estimate the metallicity [Fe/H]$_{\rm ZW}\sim$-1.68$\pm$0.10 or [Fe/H]$_{\rm
UVES}\sim$-1.64$\pm$0.13, log-luminosity $\log(L/L_{\odot})\sim$1.76$\pm$0.02, absolute magnitude M$_V\sim$0.38$\pm$0.04~mag, and true
distance modulus of $\mu_0\sim$16.93$\pm$0.04~mag, which is equivalent to a distance of $\sim$24.3$\pm$0.5~kpc. All of these values are consistent
with previous estimates in the literature.}
   {}


   \maketitle
%
%

\section{Introduction}
\label{sec:intro}

The potential of difference image analysis (DIA) as a powerful tool for unveiling short period variable stars, or small amplitude variations in
Blazhko RR Lyraes, in the densely populated central regions of globular clusters (GC), has been demonstrated in recent papers \citep[e.g.][
etc]{kains12+05, af12+04, af11+04, bramich11+03, corwin06+06, strader02+02}.
Multi-colour time-series CCD photometry allows the
identification of variable stars in specific regions of the colour-magnitude diagram (CMD). The Fourier decomposition of RR Lyrae light curves enables
us to derive stellar parameters. In the Blue Straggler region, SX Phe stars are often found and depending on the number of them
in the cluster, their Period-Luminosity relation (P-L) can be calibrated or it can be used to obtain an independent estimate of the distance to the
cluster.

In the present paper we focus on the globular cluster NGC~7492. This is a rather sparse outer-halo cluster ($R_{GC} \sim$25~kpc; \cite{harris96}
(2010 edition)) for which detailed spectroscopic abundances exist ([Fe/H] $\sim-$1.8) for four stars at the tip of the red giant branch (RGB)
\citep{cohen05+01}. Hence
the cluster offers a good opportunity to compare the spectroscopic results with the metallicity derived from the light curve Fourier decomposition
approach for RR Lyrae stars. According to the Catalogue of Variable Stars in Globular Clusters (CVSGC) \citep{clement01}, only four variable stars are
known in this cluster; namely one RR0 (V1), discovered by \cite{shapley20XVII}, two RR1 (V2 and V3) and one long period variable (LPV; V4), all
discovered by \cite{barnes68}. Although \cite{buonanno87+03} suggest the presence of a population of blue stragglers in this cluster
and numerous blue stragglers have been identified by \cite{cote91+02}, no investigation into the variability of these otherwise faint stars
($V\sim$20~mag) has been reported. Taking advantage of our time-series CCD photometry of the cluster and our capability of performing precise
photometry via DIA, we explore the field of the cluster for new variables.

In $\S$ \ref{sec:ObserRed} we describe the observations, data reduction, and transformation of the photometry to the Johnson-Kron-Cousins standard
system. In $\S$ \ref{sec:var_find_technique} we present a detailed discussion of the strategies employed for the identification of new variables. In
$\S$ \ref{sec:RRLyrae} the physical parameters of the RR0 star as derived from the Fourier decomposition of its light curve and the Blazhko effect for
the RR1 star are discussed. In $\S$ \ref{sec:newRGB}, we make brief comments on the long term variables. In $\S$ \ref{sec:newsx} we present a
discussion for the newly found SX Phe stars and candidates and in $\S$ \ref{sec:Concl} we summarize our results.

\section{Observations and reductions}
\label{sec:ObserRed}

The observations employed in the present work were obtained, using the Johnson-Kron-Cousins $V$, $R$ and $I$ filters, on the dates listed in table
\ref{tab:date}. We used the 2.0m telescope of the Indian Astronomical Observatory (IAO) at Hanle, India, located at 4500m above sea level. The typical
seeing was $\sim$1.3 arcsec. The detector was a Thompson CCD of 2048 $\times$ 2048 pixels with a pixel scale of 0.296 arcsec/pix and a field of view
of $\sim10.1 \times 10.1$ arcmin$^2$. However, for this cluster we can only apply DIA to smaller images that cover an area of $\sim$6.4$\times$5.5
arcmin$^2$ centred on the cluster because of a lack of sources towards the detector edges for use in the kernel solutions. Our data set consists of
119 images in $V$, 54 images in $R$, and 10 images in $I$.

The images were calibrated via standard overscan bias level and flat-field correction procedures, and difference image analysis (DIA) was performed
with the aim of extracting high precision time-series photometry of the stars in the field of NGC~7492. We used the {\tt DanDIA}\footnote{
{\tt DanDIA} is built from the DanIDL library of IDL routines available at http://www.danidl.co.uk} pipeline for the data reduction process which
models the convolution kernel matching the point-spread function (PSF) of a pair of images of the same field as a discrete pixel array
\citep{bramich08,Bramichetal2012}. A brief summary of the {\tt DanDIA} pipeline can be found in \cite{af11+04} while a detailed description of the
procedure and its caveats is available in \cite{bramich11+03}.

\begin{table}
\scriptsize
\caption{Distribution of observations of NGC 7492 for each filter.}
\centering
\begin{tabular}{ccccccc}
\hline
Date     &$N_V$&$t_V$(s) &$N_R$&$t_R$(s)& $N_I$ & $t_I$(s)\\
\hline
20041004 &  7  & 60-200  &  7  & 150-180&  --   &   --    \\
20041005 & 16  & 60-120  & 14  &   100  &  --   &   --    \\
20060801 &  8  &  120    &  7  &   100  &  --   &   --    \\
20070804 &  6  &  240    &  7  & 100-180&  --   &   --    \\
20070805 &  4  &  240    &  3  & 120-180&  --   &   --    \\
20070904 &  8  & 180-240 &  8  & 120-180&  --   &   --    \\
20070905 &  8  &  180    &  8  &   120  &  --   &   --    \\
20090107 &  4  &  100    & --  &   --   &  5    & 100-300 \\
20090108 &  5  &  300    & --  &   --   &  5    & 220-300 \\
20120628 & 37  & 90-180  & --  &   --   &  --   &   --    \\
20120629 & 16  &  200    & --  &   --   &  --   &   --    \\
\hline
TOTAL:   & 119 &         & 54  &        &  10   &         \\
\hline
\end{tabular}
\tablefoot{
\tablefoottext{a}{$N_V$, $N_R$ and $N_I$ are the number of images taken for the filters $V$, $R$, and $I$ respectively.}
\tablefoottext{b}{$t_V$, $t_R$ and $t_I$ are the exposure times, or range of exposure times, employed during each night for each filter.}}
\label{tab:date}
\end{table}

The reference image for each filter was constructed by registering and stacking the best-seeing calibrated images such that all images used were
taken on a single night. This resulted in 2, 4, and 1 images being stacked with total exposure times of 120, 400 and 100 s for the filters $V$, $R$
and $I$, respectively.

The light curve data in all three filters for all of the variable stars is provided in table \ref{tab:vri_phot}. In addition to the star magnitudes,
we supply the difference fluxes $f_{\mbox{\scriptsize diff}}(t)$ (ADU/s), the reference flux $f_{\mbox{\scriptsize ref}}$ (ADU/s) and the photometric
scale factor $p(t)$, at time $t$, as provided by the {\tt DanDIA} pipeline. These quantities are linked to the instrumental magnitudes
$m_{\mbox{\scriptsize ins}}$ via the equations

\begin{equation}
f_{\mbox{\scriptsize tot}}(t) = f_{\mbox{\scriptsize ref}} + \frac{f_{\mbox{\scriptsize diff}}(t)}{p(t)}
\label{eq:totflux}
\end{equation}
\begin{equation}
m_{\mbox{\scriptsize ins}}(t) = 25.0 - 2.5 \log (f_{\mbox{\scriptsize tot}}(t)).
\label{eq:mag}
\end{equation}

\subsection{Transformation to the $VRI$ standard system}
\label{subs:transf_std}
\begin{figure}[!t]
\begin{center}
\includegraphics[scale=0.47]{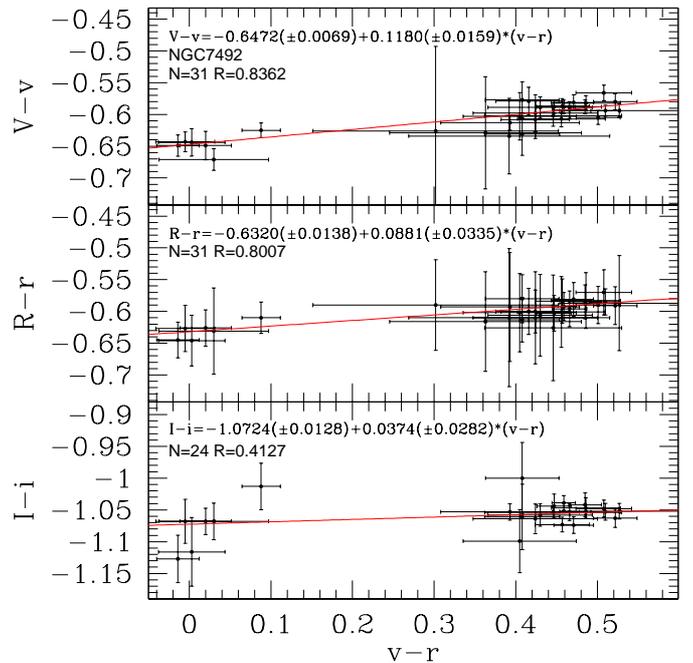}
\caption{Photometric transformation relations between the instrumental $v, r, i$ and the standard $V, R, I$ magnitudes using the standard stars
from Stetson (2000).}
    \label{trans}
\end{center}
\end{figure}

The instrumental $v$, $r$, $i$, magnitudes were converted to the Johnson-Kron-Cousins photometric system \citep{landolt92} by using the standard stars
in the field of NGC 7492. The $V$, $R$ and $I$ standard stars and their magnitudes are available in the catalogue of Stetson
(2000)\footnote{http://www3.cadc-ccda.hia-iha.nrc-cnrc.gc.ca/community/STETSON/standards}. Fig. \ref{trans} displays the relations between the
instrumental and standard magnitude systems as a function of instrumental ($v-r$) colour, where we found mild colour dependencies. The standard stars
that we used have a colour range between -0.02$<V-R<$0.53~mag and 0.02$<V-I<$1.10~mag which covers the range of colours of the stars in our field
of view. To convert the instrumental magnitudes to standard magnitudes for the stars without an instrumental $v-r$ colour, we assumed a value of $v-r
= 0.286$~mag corresponding to the centre of the spread of instrumental $v-r$ star colours.

\subsection{Astrometry and finding chart}
\label{sub:finding_chart}

\begin{figure*}[!htp]
\begin{center}
\includegraphics[scale=0.75]{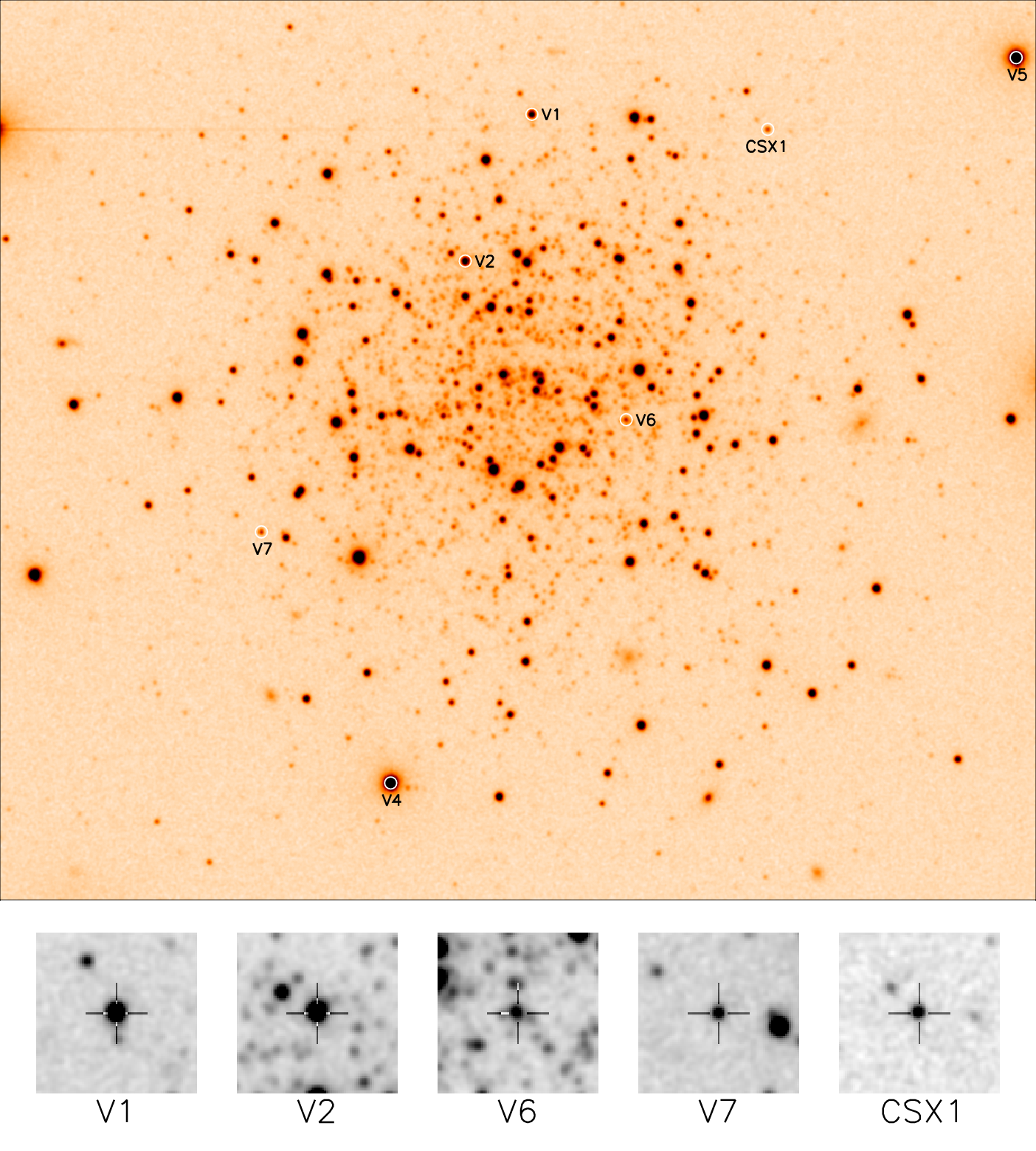}
\caption{Finding charts constructed from our $V$ reference image; north is up and east is to the right. The cluster image is 6.36$\times$5.52
arcmin$^{2}$, and the image stamps are of size 23.7$\times$23.7 arcsec$^{2}$. Each variable (except V4 and V5) lies at the centre of its corresponding
image stamp and is marked by a cross-hair.}
\label{fig:N7492id}
\end{center}
\end{figure*}

We fit a linear astrometric solution derived for the $V$ filter reference image by matching 37 hand-picked stars with the UCAC3 star
catalogue \citep{ucac3} using a field overlay in the image display tool GAIA \citep{Draper2000}. We achieved a radial RMS scatter in the residuals
of $\sim$0.24~arcsec which is equivalent to $\sim$0.82 pixels. To facilitate the identification of the variable stars in this cluster in future
studies, we have produced a finding chart which we present in figure \ref{fig:N7492id}. In addition, in Table \ref{tab:astrom} we present the
equatorial J2000 celestial coordinates of all of the variable stars discussed in this work. This astrometric solution is in perfect agreement
with the astrometry given by \cite{stetson00+00} for the standard stars in this cluster.

\begin{table*}[htp]
\caption{Celestial coordinates for all of the confirmed and candidate variables in our field of view, except V3 which lies outside of our field of
view. The coordinates correspond to the epoch of the $V$ reference image, which is the heliocentric Julian date $\sim$2453284.26~d. We also include
in this table, the epoch, period, mean $V$ magnitude, $V-R$ colour and the full amplitude of each variable.} 
\centering
\begin{tabular}{cccccccccc}
\hline

ID  &  type  &   RA        &     DEC    &      Epoch   & P                       &   $V$                   &$V-R$                 & Amp (V)  &Amp(R)\\
    &        &  (J2000)    &    (J2000) &     (days)   & (days)                  &  (mag)                  &(mag)                 & (mag)    &(mag) \\
\hline
V1  & RR0    & 23:08:26.68 &-15:34:58.5 & 2453949.4046 & 0.805012                & 17.303\tablefootmark{b} &0.264\tablefootmark{b}&0.511     &0.408\\
V2  & RR1    & 23:08:25.00 &-15:35:52.9 & 2453284.2652 &0.411764\tablefootmark{a}& 17.256\tablefootmark{b} &0.188\tablefootmark{b}&0.251     &0.140\\
V4  & LPV    & 23:08:23.19 &-15:39:06.0 & --           &$\sim$21.7               & 14.271\tablefootmark{c} & --                   &$\sim$0.18&  -- \\
V5  & LPV    & 23:08:39.08 &-15:34:36.3 & --           &--                       & 14.258\tablefootmark{c} & --                   &$>$0.33   &  --
\\
V6  & SX Phe & 23:08:29.16 &-15:36:51.1 & 2454318.3687 &0.0565500                & 19.235\tablefootmark{b} &0.166\tablefootmark{b}&0.136     &0.058\\
V7  & SX Phe & 23:08:19.83 &-15:37:33.6 & 2454839.0479 &0.0725859                & 19.363\tablefootmark{b} &0.448\tablefootmark{b}&0.050     &  -- \\
CSX1& SX Phe?& 23:08:32.73 &-15:35:03.4 & --           & --                      & 19.499\tablefootmark{c} &0.210\tablefootmark{c}&$\sim$0.15&0.09?\\
\hline
\end{tabular}
\tablefoot{\tablefoottext{a}{If we consider a secular period change then the period is P$_0=0.412119$~d at the epoch $E=$2453284.2652~d and the period
change rate is $\beta\approx$ 47~d~Myr$^{-1}$. }\tablefoottext{b}{Intensity weighted magnitude calculated from the light curve
model.}\tablefoottext{c}{Mean magnitude from our data.}}
\label{tab:astrom}
\end{table*}

\begin{table*}
\caption{Time-series $V$, $R$ and $I$ photometry for all of the confirmed and candidate variables in our field of view. Note that V3 lies outside
of our field of view. The standard $M_{\mbox{\scriptsize std}}$ and instrumental $m_{\mbox{\scriptsize ins}}$ magnitudes are listed in columns 4 and
5, respectively, corresponding to the variable star, filter, and heliocentric Julian Date of mid-exposure listed in columns 1-3,
respectively. The
uncertainty on $m_{\mbox{\scriptsize ins}}$ is listed in column 6, which also corresponds to the uncertainty on $M_{\mbox{\scriptsize std}}$. For
completeness, we also list the quantities $f_{\mbox{\scriptsize ref}}$, $f_{\mbox{\scriptsize diff}}$ and $p$ from Equation~\ref{eq:totflux} in
columns 7, 9 and 11, along with the uncertainties $\sigma_{\mbox{\scriptsize ref}}$ and $\sigma_{\mbox{\scriptsize diff}}$ in columns 8 and 10. This
is an extract from the full table, which is available with the electronic version of the article.
         }
\centering
\begin{tabular}{ccccccccccc}
\hline
Variable & Filter & HJD & $M_{\mbox{\scriptsize std}}$ & $m_{\mbox{\scriptsize ins}}$ & $\sigma_{m}$ & $f_{\mbox{\scriptsize ref}}$ &
$\sigma_{\mbox{\scriptsize ref}}$ & $f_{\mbox{\scriptsize diff}}$ &
$\sigma_{\mbox{\scriptsize diff}}$ & $p$ \\
Star ID  &        & (d) & (mag)                        & (mag)                        & (mag)        & (ADU s$^{-1}$)               & (ADU s$^{-1}$)  
                 & (ADU s$^{-1}$)                &
(ADU s$^{-1}$)                     &     \\
\hline
V1 & $V$ & 2453283.25245    & 17.400 & 18.014 & 0.009  & 558.913& 0.904 & 64.366& 4.992& 1.0059 \\
V1 & $V$ & 2453283.28310    & 17.444 & 18.058 & 0.005  & 558.913& 0.904 & 39.065& 2.618& 0.9944 \\
\vdots   & \vdots & \vdots  & \vdots & \vdots & \vdots & \vdots & \vdots& \vdots&\vdots& \vdots \\
V1 & $R$ & 2453283.27903    & 17.162 & 17.769 & 0.004  & 786.605& 0.821 & -5.906 & 3.044& 1.0210 \\
V1 & $R$ & 2453283.28689    & 17.153 & 17.759 & 0.004  & 786.605& 0.821 & 0.832 & 3.034& 1.0184 \\
\vdots   & \vdots & \vdots  & \vdots & \vdots & \vdots & \vdots & \vdots& \vdots&\vdots& \vdots \\
V1 & $I$ & 2454839.03764    & 16.575 & 17.636 & 0.010  & 865.889& 3.651 & 16.109& 8.462& 0.9974 \\
V1 & $I$ & 2454839.04150    & 16.577 & 17.639 & 0.007  & 865.889& 3.651 & 14.141& 5.362& 0.9945 \\
\vdots   & \vdots & \vdots  & \vdots & \vdots & \vdots & \vdots & \vdots& \vdots&\vdots& \vdots \\
\hline
\end{tabular}
\label{tab:vri_phot}
\end{table*}

\section{Variable star search strategies}
\label{sec:var_find_technique}

\subsection{Standard deviation}
\label{subs:SD}
\begin{figure}[!t]
\begin{center}
\includegraphics[scale=0.485]{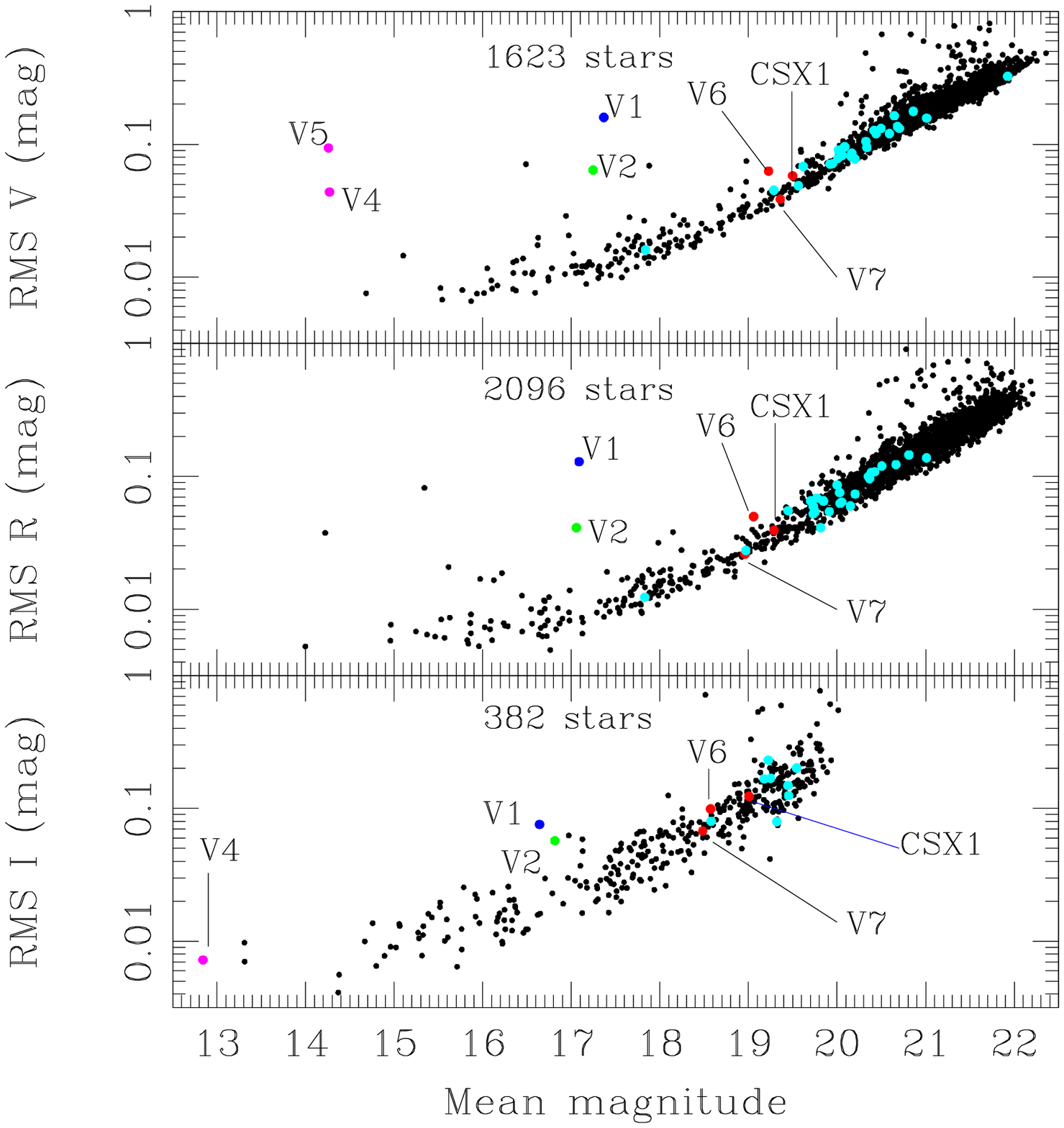}
\caption{RMS magnitude deviation as a function of the mean magnitude in the filters $V$, $R$, and $I$. Known variables are labelled. V1 (dark blue
point) and V2 (green point) are two known RR Lyrae stars in the field of the cluster. Red circles correspond to two newly identified SX Phe stars and
one candidate SX Phe. The variable V4 is saturated in our $R$ reference image and the newly identified variable V5 is saturated in the $R$ and $I$
reference images. Hence these stars are not shown in the corresponding plots. The cyan points correspond to the blue stragglers identified by
\cite{cote91+02}.}
    \label{fig:rms}
\end{center}
\end{figure}

$V$ light curves were produced for 1623 stars in the field of our images. The mean magnitude, computed using inverse variance weights, and the RMS
were calculated for each light curve. Fig. \ref{fig:rms} shows the RMS as a function of the mean magnitude for the $V$, $R$ and $I$ filters and
indicates the precision of our photometry. Stars with a large dispersion for a given mean magnitude, in principle, are good candidate variables.
However, it is possible that a light curve has a large RMS due to occasional bad measurements of the corresponding star in some images, in which case
the variability could be spurious.

We have used the RMS values as a guide to our search for variables. The two known RR Lyrae stars (V1 and V2) and the known red giant variable (V4) are
highlighted with colours as indicated in the caption of Fig. \ref{fig:rms}. They clearly stand out from the general trend. V3 is not in the field of
our images. With this method we have also identified another long period variable discussed later in this paper which we have labelled as V5.
While this method is useful for detecting bright variables, it is not useful for detecting shorter period faint variables in the blue straggler
region. It is clear from the red points, which correspond to two new SX Phe and one candidate SX Phe, that they do not stand out in the plot
relative to other faint non-variable stars. For these variables, we have used a different approach described in the following sections.

\subsection{String-length period search}
\label{subs:SQ}
\begin{figure}[!ht]
\begin{center}
\includegraphics[scale=0.475]{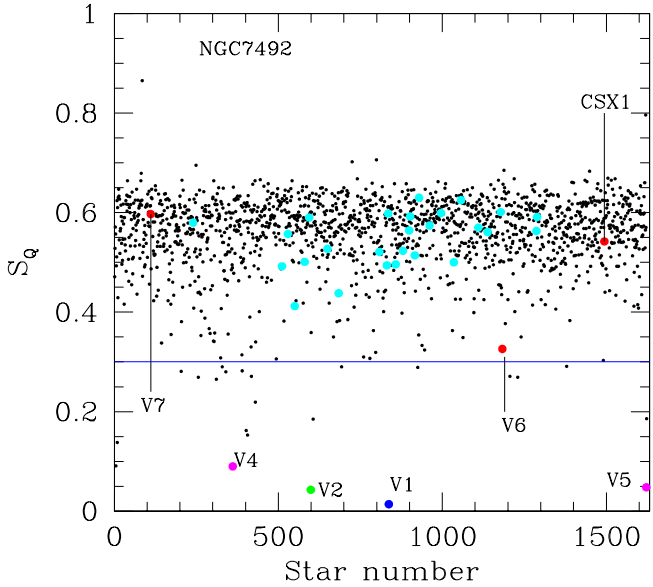}
\caption{$S_Q$ values for all the stars in the $V$ images as a function of an arbitrary star number. The blue line is the threshold below which RR
Lyrae stars tend to be found. The known variables V1, V2 and V4 are labelled as well as the newly identified long period variable V5, the new
SX Phe V6 and V7 and one candidate SX Phe star (red points). The cyan points correspond to the stars identified by \cite{cote91+02} as blue
stragglers.}
  \label{fig:SQ}
\end{center}
\end{figure}

The light curves of the 1623 stars measured in each of the 119 $V$ images were analyzed by the string-length minimization approach
\citep{burke70+02,Dworetsky83}. In this analysis, the light curve is phased with numerous test periods within a given range. For each period the
dispersion parameter $S_Q$ is calculated. When $S_Q$ is at a minimum, the corresponding period produces a phased light curve with a minimum possible
dispersion and it is adopted as the best-fit period for that light curve. Bona fide variable stars should have a value of $S_Q$ below a certain
threshold. Similar analyses of clusters with numerous variables have shown that all periodic variables with long periods and large amplitudes are
likely to have $S_Q \leq 0.3$. However short-period small-amplitude variables like SX Phe are often missed by this approach \citep{af04+05,af06+02}.
Fig. \ref{fig:SQ} shows the distribution of $S_Q$ values for all stars measured in the $V$ images, plotted as a function of an arbitrary star number.
We explored individually the light curves below the indicated threshold of $S_Q=0.3$. With this method we recover the two RR Lyrae stars in the field
of our images (V1 and V2), the LPV V4, and we discovered the LPV V5. The others stars below this threshold were not found to display true variations.

\subsection{Colour-magnitude diagram}
\label{subs:CMD}
\begin{figure}[!ht]
\begin{center}
\includegraphics[scale=0.495]{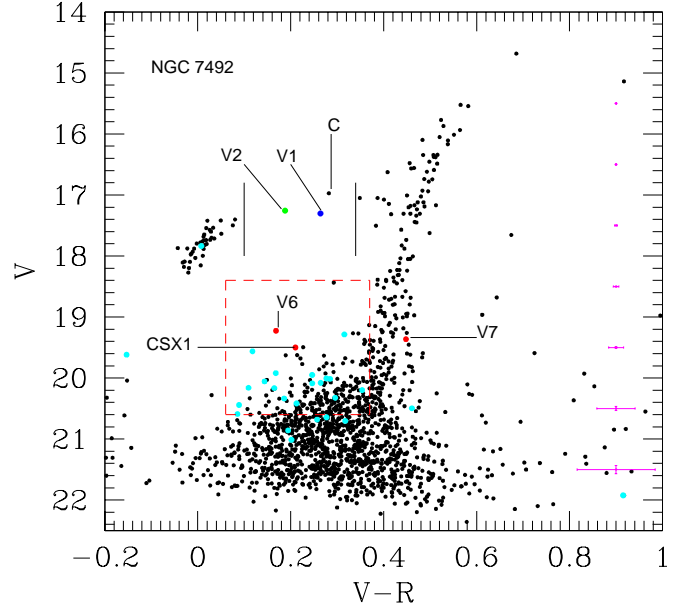}
\caption{CMD of NGC 7492. Two known RR Lyrae variables V1 and V2 are marked with dark blue and green symbols respectively. The new SX Phe variables V6
and V7 and one candidate SX Phe are shown with red symbols. The cyan points correspond to the blue stragglers identified by \cite{cote91+02}. The red
box is an arbitrarily defined Blue Straggler region (see text).}
    \label{fig:CMD}
\end{center}
\end{figure}

The CMD is very useful for separating groups of stars that are potential variables, e.g. in the horizontal branch (HB), the RGB and the blue
straggler region. Fig. \ref{fig:CMD} shows the $V$ versus $(V-R)$ diagram. The known RR0 and RR1 stars contained in our field of view are
shown as dark blue and green circles, respectively. The known red giant variable V4 is not shown because it is saturated in our $R$ reference image
and V5 is saturated in the $R$ and $I$ reference images. The blue straggler region has been arbitrarily defined by the red box in this CMD. We
selected the faint limit such that the photometric uncertainty is below 0.1 mag and the red limit so that the region is not too contaminated by the
main sequence. 

It is worth noting that in the HB, the RR Lyrae region is populated only by two of the three already known RR Lyrae stars and one more
star labelled as C in the figure. The light curve of star C does not show signs of variation at the precision of our data ($\sim$0.02~mag),
hence the star may be a field object. Furthermore, there are no saturated stars in the field of view of our $V$ filter images. Therefore, with a
typical precision of $\sim$0.01-0.02~mag in our $V$ light curves at the magnitude of the HB ($\sim$17.3~mag), we can be sure that there are no more
RR Lyrae stars in the cluster in our field of view (Fig. \ref{fig:N7492id}). 

\subsection{Variability detection statistic $\cal S_B$}
\label{subs:SB}

We also analysed the light curves for variability via the detection statistic $\cal S_B$
defined by \cite{af12+04} and employed by these authors to
detect amplitude modulations in RR Lyrae stars attributed to
the Blazhko effect. The variability detection statistic $\cal S_B$ was inspired by the alarm statistic
$\cal A$ defined by \cite{tamuz06+02}, designed originally for improving the fitting of
eclipsing binary light curves. The advantages of redefining the alarm statistic as:

\begin{equation}
S_B=\left(\frac{1}{NM}\right)\sum_{i=1}^{M}\left(\frac{r_{i,1}}{\sigma_{i,1}}
+\frac{r_{i,2}}{\sigma_{i,2}}+...+\frac{r_{i,k_{i}}}{\sigma_{i,k_{i}}}
\right)^2,
\label{eq:fourier}
\end{equation}

\noindent
have been discussed by \cite{af12+04}. In this equation, $N$ represents the total
number of data points in the light curve and $M$ is the number of groups of
time-consecutive residuals of the same sign from a constant-brightness light curve model. The
residuals $r_{i,1}$ to $r_{i,k_i}$ form the $i$th group of $k_i$ time-consecutive
residuals of the same sign with corresponding uncertainties $\sigma_{i,1}$ to $\sigma_{i,k_i}$.
Our $\cal S_B$ statistic may therefore be interpreted as a measure of the systematic
deviation per data point of the light curve from a non-variable (constant-brightness) model.
We note that in \cite{af12+04} the residuals $r_{i,j}$ are calculated relative to the
Fourier decomposition light curve model rather than relative to a constant-brightness model used in this
work. It is this difference in application that makes $\cal S_B$ a detection statistic for the Blazhko effect in \cite{af12+04} and a
detection statistic for variability in this paper.

Equation~\ref{eq:fourier} has been modified from the corresponding equation
in \cite{af12+04} by further normalising the $\cal S_B$ statistic by $M$. This modification
serves to improve the discriminative power of the statistic because variable stars,
as opposed to non-variable stars, have longer time-consecutive runs of light curve
data points that are brighter or fainter than the constant-brightness
model, and therefore smaller values of $M$ (for a given light curve $N$).

We calculated $\cal S_B$ for each of our $V$ and $R$ light curves and we made plots of
$\cal S_B$ versus magnitude in each filter. The variables detected so far by the methods
discussed in Sections \ref{subs:SD}-\ref{subs:CMD} (V1, V2, V4 and V5) stand out in these diagrams with very large $\cal S_B$
values compared to the other stars. However, we found that we could make these differences
in the $\cal S_B$ values between variable and non-variable stars even larger by calculating
$\cal S_B$ for the combined $VR$ light curves. In this case, for each star, we adjust the 
$R$ light curve so that its mean magnitude matches that of the corresponding $V$ light curve, and then
we calculate $\cal S_B$ for the combined $VR$ light curve\footnote{This procedure is valid for the
variable stars in our data because the data points in our light curves generally alternate between
the two filters and therefore the light curve data in each filter has approximately the same phase coverage.}.

\begin{figure}[!t]
\begin{center}
\includegraphics[scale=0.48]{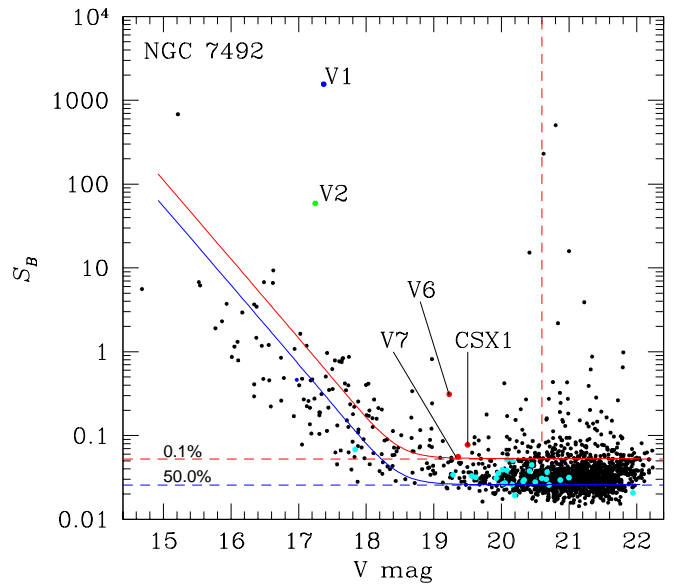}
\caption{$\cal S_B$ statistic as a function of $V$ mean magnitude for the $VR$ combined
         light curves. The RR Lyrae stars V1 and V2 are labelled, as are the new SX Phe
         stars V6 and V7 and the one candidate SX Phe star. The cyan points correspond to the blue stragglers identified by \cite{cote91+02}.
The long period
         variables V4 and V5 do not appear on this plot because they are saturated in the $R$ filter.
         The solid blue curve is the median (50\%) curve determined from our simulations and adjusted to fit
         the real $\cal S_B$ data above $V \sim$19~mag. The dashed red curve represents our variable star detection
         threshold in $\cal S_B$ set using our simulations to limit our false alarm rate to $\sim$0.1\%. The solid red curve represents our adopted
variable star detection threshold when we take into account the systematic errors. We further limited our variable star search to stars brighter than
$V =$20.6~mag (vertical dashed red line).}
\label{fig:alarm}
\end{center}
\end{figure}

In Fig. \ref{fig:alarm} we plot $\cal S_B$ for each of the combined $VR$ light curves as a function
of the $V$ mean magnitude. The RR Lyrae stars V1 and V2 clearly have $\cal S_B$ values among the largest in the
light curve sample. It is interesting to note that $\cal S_B$ generally scatters around a constant value ($\sim$0.03)
for $V$ fainter than $\sim$19~mag. For stars brighter than $V \sim$19~mag, the $\cal S_B$ values show an exponential increase
(which appears as linear on the log-scale of Fig. \ref{fig:alarm}). This feature can be explained by considering the systematic errors
that exist at some level in all the light curves. However, we defer the relevant dicussion of this topic until later in this section.

To detect new variable stars, we need to define a detection threshold that optimises our sensitivity to real variables while
being set high enough to minimise the number of false alarms (i.e. classification of non-variable stars as variable). 
Without setting this detection threshold carefully, one runs the risk of publishing suspected variable stars of which the
majority may be refuted in subsequent photometric campaigns (see \cite{safonova11+01} and \cite{bramich12+03} papers for a good example).
We decided to determine the threshold for our $\cal S_B$ statistic through the use of simulations.

For each combined $VR$ light curve in our sample, we performed 10$^{6}$ simulations. Each simulation consists of generating
a random light curve $m_{i}$ using the real light curve data
point uncertainties $\sigma_{i}$ via:
\begin{equation}
m_{i} = \overline{V} + \lambda_{i} \sigma_{i},
\label{eqn:randomlc}
\end{equation}
where the $\lambda_{i}$ are a set of random deviates drawn from a normal distribution with zero mean and unit $\sigma$,
and $\overline{V}$ is the mean $V$ magnitude of the real light curve. We calculated $\cal S_B$ for the simulated
light curves and obtained a distribution of 10$^{6}$ $\cal S_B$ values from which we determined the median (50\%)
and 99.9\% percentile.

We found that the median values of the $\cal S_B$ distributions are approximately the same (to within the noise of the finite
number of simulations) for all of our stars, as are the 99.9\% percentile values, which implies that for light curves with the same number of data
points, the actual distribution of data point uncertainties has no impact on the threshold to be chosen for $\cal S_B$. We found that for our combined
$VR$ light curves, the mean of the $\cal S_B$ distribution medians is $\sim$0.0256, and the mean of the 99.9\% percentiles is $\sim$0.0525. These
lines are plotted in Fig. \ref{fig:alarm} as the horizontal dashed blue and red lines.

Looking again at Fig. \ref{fig:alarm} we now see that the $\cal S_B$ values for the real light curves scatter close to the median line
from the
simulations for stars fainter than $V \sim$19~mag, which implies that for these stars the simulations provide a reasonably good model for the
noise in the real
light curves. However, for the stars brighter than $V \sim$19~mag, the $\cal S_B$ values for the real light curves increase exponentially with
increasing brightness and are much larger than
what we would expect as determined from our light curve simulations with pure Gaussian noise. We can explain this by considering
that the systematic errors in the light curves, which correlate over groups of time-consecutive data points and therefore mimic
real variability, increasingly dominate the noise in our real light curves with increasing star brightness. To account for the systematic errors, we
need to adjust our median
and 99.9\% percentile curves in Fig. \ref{fig:alarm}, which we do by fitting a linear relation to the log-$\cal S_B$ values for $V$ brighter than
19~mag and merging this fit with the constant median curve for $V$ fainter than 19~mag (solid blue curve). We then shift this curve to larger $\cal
S_B$
values so that the horizontal part matches that of the 99.9\% percentile (solid red curve). Finally we adopt the solid red curve as our detection
threshold
for new variables.

By choosing a variable star detection threshold set to the 99.9\% percentile of $\cal S_B$ from our simulations of light curves that have only pure
Gaussian noise,
we have set our false alarm rate to 0.1\%, which implies that with 1585 stars with combined $VR$ light curves we should expect only $\sim$1.6
non-variable stars to fall
above our threshold. However, since we are fully aware that the systematic errors may affect some light curves more than others for various reasons
(e.g. near a saturated
star, cosmic ray hits, etc.) we must still exercise caution with all candidate variable stars that lie above our detection threshold in $\cal S_B$.
We observe that for stars fainter than $V =$20.6~mag, the $\cal S_B$ values have a larger number of high outliers than is typical and therefore we
further
limit our variable star search to stars brighter than $V =$20.6~mag (vertical dashed red line in Fig. \ref{fig:alarm}). We note that the two RR Lyrae
stars V1 and V2 have
$\cal S_B$ values much greater than our adopted detection threshold and are therefore recovered by this method.

We have explored the appearance of the light curves of all stars with $\cal S_B$ above our detection threshold and we have found convincing
indications of variability in two stars in the blue straggler region. These SX Phe stars are discussed in Section~\ref{sec:newsx} along with one other
candidate SX Phe star that also lies above our detection threshold in $\cal S_B$. The remaining stars with $\cal S_B$ values above our threshold do
not show convincing light curve variability either on inspection of their light curves or when analysed with the string-length minimisation approach.

If we compare this method with the others used in this paper for detecting variable stars (see sec. \ref{subs:SD}, \ref{subs:SQ}), then it
becomes clear that this method is the only one that has been used to successfully detect all of the previously known and new variables in this
cluster.

\section{The RR Lyrae stars}
\label{sec:RRLyrae}

Of the three known RR Lyrae stars in the cluster, the RR1 star V3 is not in the field of our images. 

\begin{figure}
\begin{center}
\includegraphics[scale=0.51]{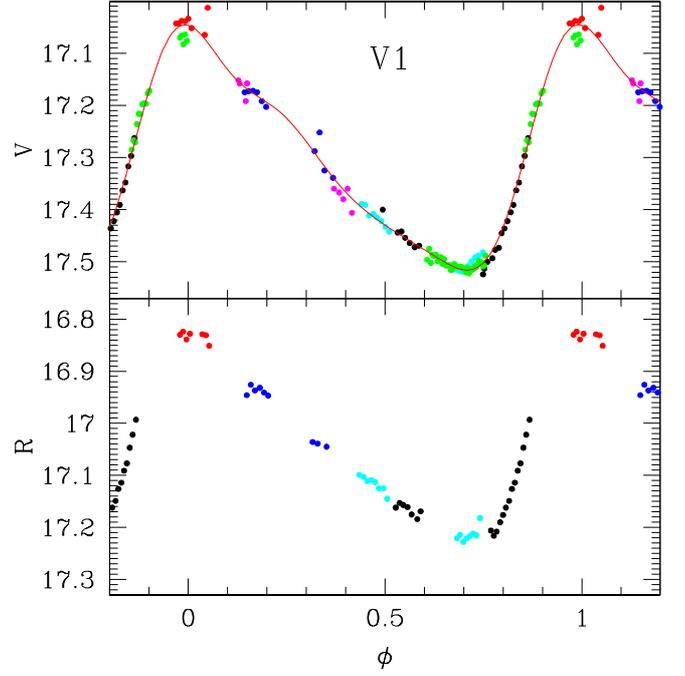}
\caption{Light curve of the RR0 star V1 in the $V$ filter (top) and the $R$ filter (bottom) phased with the period 0.805012~d. The data point colours represent the different epochs listed in table \ref{tab:colour_code}. The red line
corresponds to the Fourier fit of equation \ref{eq:fourier} with four harmonics. The typical uncertainties in the $V$ and $R$ magnitudes are
$\sim$0.007 and 0.005~mag, respectively.}
\label{fig:V1}
\end{center}
\end{figure}

{\bf V1}. This is a clear fundamental mode pulsator or RR0. Our data are neatly phased with a period of 0.805012~d (Fig \ref{fig:V1}). The light curve
was fitted with 4 Fourier harmonics (red continuous line in Fig. \ref{fig:V1}) of the form of eq. \ref {FOUfit}.

\begin{equation}\label{FOUfit}
m(t) = A_o ~+~ \sum_{k=1}^{N}{A_k ~cos~( {2\pi \over P}~k~(t-E) ~+~ \phi_k ) }.
\end{equation}

We noted that using more than four harmonics results in over-fitting. The decomposition of the light curve in Fourier harmonics was used to
estimate the iron abundance [Fe/H] and the absolute magnitude $M_V$, and hence
the distance. These calculations were made using the semi-empirical calibrations available in the literature.

To calculate [Fe/H] we employed the calibration of \cite{jurcsik96+01} valid for RR0 stars. 

\begin{equation}\label{eq:JK96}
	{\rm [Fe/H]}_{J} = -5.038 ~-~ 5.394~P ~+~ 1.345~\phi^{(s)}_{31}.
\end{equation}

The Fourier parameter $\phi^{(s)}_{31}$ comes from fitting a sine series to the light curve of star V1 rather than a cosine series as in
eq. \ref{FOUfit}. However, the corresponding cosine parameter $\phi^{(c)}_{31}$ is related by  $\phi^{(s)}_{31} = \phi^{(c)}_{31} - \pi$.  The above
equation gives [Fe/H]$_J$ with a standard deviation from this calibration of 0.14 dex \citep{jurcsik98}. The Jurcsik metallicity scale  can
be transformed to the \cite{zinn84+01} scale [Fe/H]$_{ZW}$ through the relation [Fe/H]$_{\rm J}$ = 1.43 [Fe/H]$_{ZW}$ + 0.88
\citep{Jurcsik95}. We note that the deviation parameter $Dm$ \citep{jurcsik96+01} for this star when fitting higher order Fourier series is
greater than the recommended value. Hence, our metallicity estimate should be treated with caution.

We also calculate the metallicity on the UVES scale using the equation of \cite{carretta09+04}:

\begin{equation}\label{eq:Fe.uves.carretta09}
	{\rm [Fe/H]}_{\rm{UVES}} = -0.413 +0.130\rm{[Fe/H]_{ZW}}-0.356\rm{[Fe/H]}^2_{\rm{ZW}}.
\end{equation}

From our light curve fit, we find $\phi^{(c)}_{31}=$8.987 and obtain [Fe/H]$_{\rm{ZW}}=-1.68\pm$0.10 or [Fe/H]$_{\rm{UVES}}=$-1.64$\pm$0.13.
These metallicity values are in good agreement with the mean spectroscopic values of [Fe/H]=-1.82$\pm$0.05 and [Fe/H]=-1.79$\pm$0.06 determined by
\cite{cohen05+01} from Fe~I and Fe~II lines, respectively, in four bright red giants in the cluster. The Fe abundances were derived using high
resolution (R=$\lambda/\delta\lambda$=35,000) spectra obtained with HIRES at the Keck Observatory. Similarly they are in good agreement with
the latest spectroscopic metallicities from \cite{saviane12+06} (see table \ref{tab:Fe.values}).

Other metallicity estimates for this cluster are [Fe/H]=-1.70$\pm$0.06 from \cite{rutledge97+02} determined using moderate dispersion spectroscopy
in the region of the infrared Ca triplet, [Fe/H]=-1.5$\pm$0.3 from \cite{zinn84+01} using the narrow band Q39 photometric system, and
[Fe/H]=-1.34$\pm$0.25 by \cite{smith84} using the $\Delta$S method for two RR Lyraes in the cluster. Thus the metallicity estimated via the Fourier
decomposition technique agrees well with the other independent estimates (see table \ref{tab:Fe.values}).

\begin{table*}
\caption{Metallicity estimates for NGC 7492 on the ZW scale and their respective values on the UVES scale and vice versa as found from the literature
search.}
\centering
\begin{tabular}{cccc}
\hline
[Fe/H]$_{\rm{ZW}}$  &    [Fe/H]$_{\rm{UVES}}$    & Reference           &  Method \\
\hline
-1.68$\pm$0.10 &-1.64$\pm$0.13\tablefootmark{c}         & This work           & Fourier light-curve decomposition of the RR Lyrae stars\\
               &-1.72$\pm$0.07         &\cite{saviane12+06}  & Ca$_{\rm{II}}$ triplet using the FORS2 imager and spectrograph at the VLT\\
               &-1.69$\pm$0.08         &\cite{saviane12+06}  & Ca$_{\rm{II}}$ triplet using the FORS2 imager and spectrograph at the VLT\\
               &-1.69$\pm$0.08         & \cite{carretta09+04}& Weighted average of several metallicities\tablefootmark{b}\\
-1.82$\pm$0.05 &-1.83$\pm$0.07\tablefootmark{c}         &\cite{cohen05+01}    & Fe$_{\rm{I}}$ line in bright red giants in this cluster\\
-1.79$\pm$0.06 &-1.79$\pm$0.08\tablefootmark{c}         &\cite{cohen05+01}    & Fe$_{\rm{II}}$ line in bright red giants in this cluster\\
-1.70$\pm$0.06 &-1.66$\pm$0.08\tablefootmark{c}         &\cite{rutledge97+02} & Infrared Ca triplet\\
               &-1.78                  & \cite{harris96}     &Globular cluster catalogue\tablefootmark{a}\\
-1.5$\pm$0.3   &-1.41$\pm$0.36\tablefootmark{c}         &  \cite{zinn84+01}   & Narrow band Q39 photometric system\\
-1.34$\pm$0.25 &-1.23$\pm$0.27\tablefootmark{c}         &  \cite{smith84}     & $\Delta$S method for two RR Lyraes in the cluster\\
\hline
\hline
\end{tabular}
\tablefoot{\tablefoottext{a}{The catalogue version used is the updated 2010 version available at http://www.physics.mcmaster.ca/Globular.html.}
\tablefoottext{b}{\cite{carretta09+04}, \cite{carretta97+01}, \cite{kraft03+01}, and the recalibration of the Q39 and W$^{''}$ indices.}
\tablefoottext{c}{Converted from column 1 using Eq. \ref{eq:Fe.uves.carretta09} \citep{carretta09+04}.}}
\label{tab:Fe.values}
\end{table*}

For the determination of the absolute magnitude of V1 we employed the calibration of \cite{kovacs01+01},

\begin{equation}\label{eq:KW01}
M_V = ~-1.876~\log~P ~-1.158~A_1 ~+0.821~A_3 + K,
\end{equation}

\noindent
which has a standard deviation of 0.04 mag. From our fit of equation \ref{FOUfit} to the light curve of V1, we derive $A_1$=0.206 and $A_3$=0.034
mag. We adopt $K$=0.41 mag in order to be consistent with a true distance modulus for the Large Magellanic Cloud (LMC) of
$\mu_0$=18.5 mag \citep[see the discussion in][ in their section 4.2]{af10+02}. We obtain $M_V=$0.376$\pm$0.040 mag which is equivalent to the
luminosity $\log(L/L_{\odot})=$1.762$\pm$0.016. Assuming $E(B-V)=$0.0 mag \citep{harris96}, the true distance modulus is $\mu_0=$16.927$\pm$0.040
mag, equivalent to a distance of $24.3\pm0.5$~kpc. \cite{cote91+02} estimated a distance of 26.18$\pm$2.41~kpc to NGC 7492 using the cluster NGC 6752
as a reference and whose distance was estimated by \cite{PD86}. Our distance estimate using V1 agrees within the uncertainties.

\begin{figure}[!t]
\begin{center}
\includegraphics[scale=0.51]{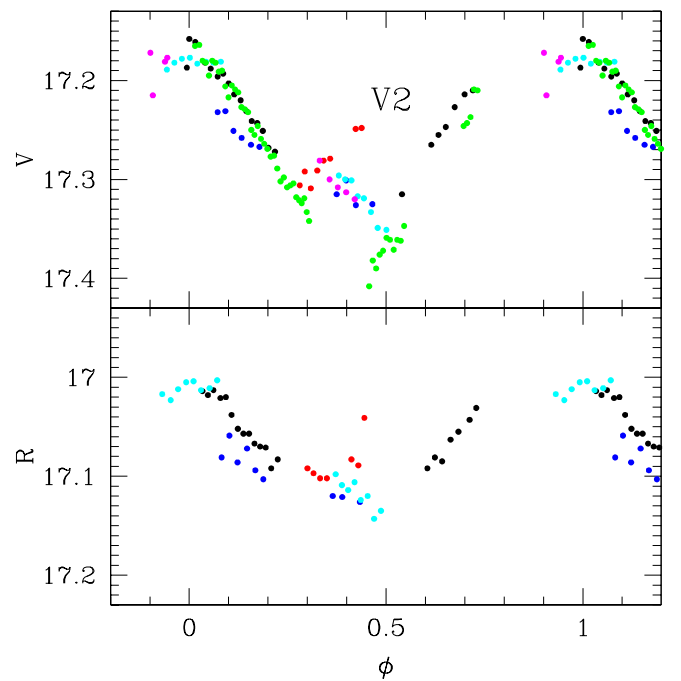}
\caption{Light curve of the RR1 star V2 in the $V$ filter (top) and the $R$ filter (bottom) phased with the period 0.411764~d. The data point colours represent the different epochs listed in table \ref{tab:colour_code}. The typical uncertainties in the $V$ and $R$ magnitudes are
$\sim$0.007 and 0.005~mag, respectively.}

\label{fig:V2}
\end{center}
\end{figure}

{\bf V2}. This RR1 star shows a complex light curve. The period derived by \cite{barnes68} of 0.292045~d fails to phase our light curve properly.
Using the string-length minimisation method on our light curve, we determine a period of 0.411764~d, which produces the phased light curve shown in
Fig. \ref{fig:V2}. The period is towards the upper limit for an RR1 type star. However, long periods like this are not uncommon in
Oosterhoff type II clusters, which typically have similar metallicity and horizontal branch
morphology as NGC~7492 \citep{lee90+00,clement01}. We note that our light curve does not phase well at this period and so we searched for a
second
period. As the ratio $P_1/P_0=$0.746$\pm$0.001 \citep{catelan09+00, cox83+02} in RRd stars, then we expect the second
period to be $\sim$0.5519~d when performing a search of the residuals from the first period. We could only find a non-significant second period
with $P_1$=0.4365~d. The light curve shows nightly amplitude changes that resemble those found by \cite{af12+04} in the majority of RR1 stars in
NGC~5024.

We have attempted to model the light curve with a secular period change which we have parameterised as in \cite{bramich11+03}, i.e.

\begin{equation}
\label{eq:chanceperiod}
 \phi(t)=\frac{t-E}{P(t)}-\left\lfloor \frac{t-E}{P(t)}\right\rfloor
\end{equation}

\begin{equation}
 P(t)=P_0+\beta(t-E),
\end{equation}
where $\phi$(t) is the phase at time $t$, $P(t)$ is the period at time $t$, $P_0$ is the period at the epoch $E$, and $\beta$ is the rate of period
change. We searched the parameter space at fixed epoch $E$, whose value is arbitrary, for the best-fitting values of $P_0$ and $\beta$, using as a
criterion the minimum string-length statistic of the light curve. The search is done in a small range of periods around the previously determined
best-fitting period. This is the only type of period change that we can consider modelling given our limited photometric data.

\begin{figure}[!t]
\begin{center}
\includegraphics[scale=0.51]{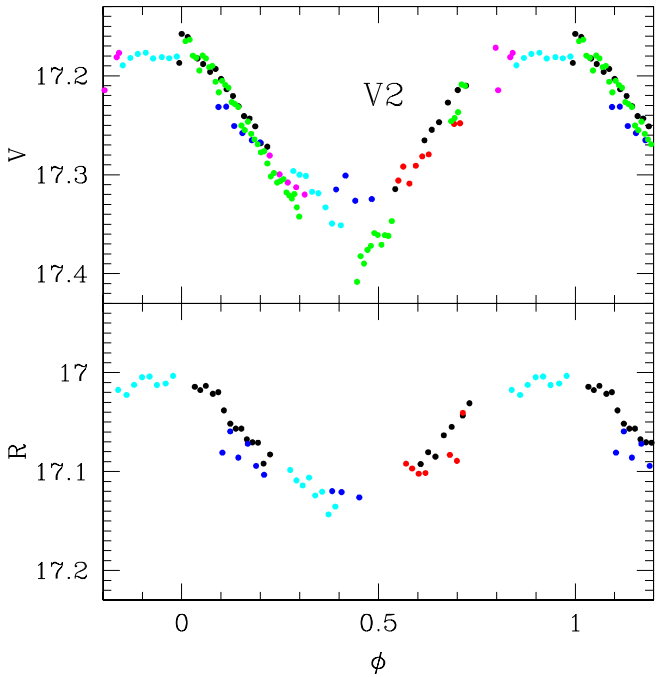}
\caption{ Same as figure \ref{fig:V2} except that now the light curve of V2 is phased with the period P$_0$=0.412119~d at the epoch $E=$2453284.2652~d
with an ephemeris that includes a period change rate $\beta\approx$47~d~Myr$^{-1}$.}
\label{fig:V2_PCH}
\end{center}
\end{figure}

We found a period of P$_0=$ 0.412119~d at the epoch $E=$2453284.2652~d and a period change rate of $\beta\approx$47~d~Myr$^{-1}$. The light
curve phased with $\phi$(t) from eq. \ref{eq:chanceperiod} is shown in Fig. \ref{fig:V2_PCH}. Clearly the phased light curve is now much improved, but
we still observe possible amplitude modulations, which may be due to the Blazhko effect. This period change rate
is higher than other values found in the literature by a factor of two or more \citep{leborgne07+08, lee91+00, jurcsik01+03}, but given our limited
data we cannot speculate on the cause.

\begin{table}
\caption{The data point colours used to mark different observing runs in figs \ref{fig:V1}, \ref{fig:V2}, \ref{fig:V2_PCH}, \ref{fig:V6} and
\ref{fig:V7}.}
\centering
\begin{tabular}{cc}
\hline
Dates   &  Colour \\
\hline
20041004 - 20041005 & Black  \\
20060801            & Red    \\
20070804 - 20070805 & Blue   \\
20070904 - 20070905 & Cyan   \\
20090107 - 20090108 & Magenta\\
20120628 - 20120629 & Green  \\
\hline
\end{tabular}
\label{tab:colour_code}
\end{table}

\section{Long period variables}
\label{sec:newRGB}

\begin{figure}[htp]
\begin{center}
\includegraphics[scale=0.51]{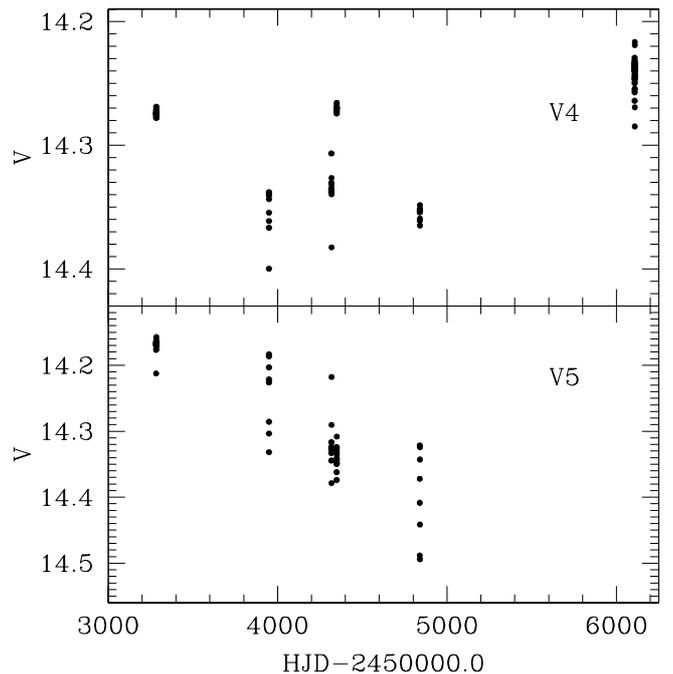}
\caption{Light curves of the variables V4 and V5 in the $V$ filter.}
\label{fig:RDG_VAR}
\end{center}
\end{figure}

{\bf V4}. This red giant variable, discovered by \cite{barnes68}, clearly stands out as a variable in Figs. \ref{fig:rms} and \ref{fig:SQ}.
\cite{barnes68} estimated a period of 17.9 days but pointed out that the observations did not cover the whole period. Our data set for this star
consists of 119 $V$ filter epochs distributed over a baseline of 8 years (top panel of Fig. \ref{fig:RDG_VAR}). Thus our data are less than ideal for
estimating an accurate period. Nevertheless, using the {\tt Period04} program \citep{period04}, we find a period of $\sim$21.7~d. In the V vs
(V-I) diagram (not plotted in the paper), the star is situated in the upper region of the red giant branch. Exploring the Catalogue of Variable Stars
in Galactic Globular Clusters \citep{clement01} one finds LBs (slow irregular variables of types K, M, C and S; see the General Catalog of Variable
Stars \citep{kholopov96+09} for classifications of variables) with periods of 13-20~d and amplitudes 0.1-0.4 mag. See for example V8 and V10 in NGC
2419. See also, V109 in NGC 5024 listed as a semi-regular variable with a period of 21.93~d and amplitude of 0.05 mag. Our data for V4 are consistent
with the classification as a long period variable.

{\bf V5}. From Figs. \ref{fig:rms} and \ref{fig:SQ} we discover this new long period variable. Its light curve is shown in the bottom
panel of Fig. \ref{fig:RDG_VAR}. This star is saturated in our $R$ and $I$ images and hence we have not been able to plot it in the CMD and determine
its classification (e.g. as a red giant). It is evident from the light curve that the star undergoes a long term dimming.

We note that for both V4 and V5, the formal uncertainties on the data points are typically $\sim$0.001 mag. However, these stars are very bright and
their light curves suffer from systematic errors that correlate during the nightly observations, leading to a relatively large intra-night
scatter (see Fig. \ref{fig:RDG_VAR}).

\section{SX Phoenicis stars and candidates}
\label{sec:newsx}

We have discovered two new SX Phe stars which we label V6 and V7, and one candidate SX Phe star which we label CSX1. 

{\bf V6}. This variable was found above our detection threshold for the $\cal S_B$ statistic in section \ref{subs:SB} (see figure \ref{fig:alarm}). In
the CMD it is placed well inside the blue straggler region (see figure \ref{fig:CMD}). In fact, this star is a blue straggler as found by
\cite{cote91+02}. We analysed the $V$ light curve with {\tt Period04} and found a clear frequency at 17.683477~cycles~d$^{-1}$ (or a period of
0.0565500~d). We did not find any further significant frequencies. Based on the blue straggler status and the detected period, we can be sure that
this is an SX Phe star.

\begin{figure}[!t]
\begin{center}
\includegraphics[scale=0.51]{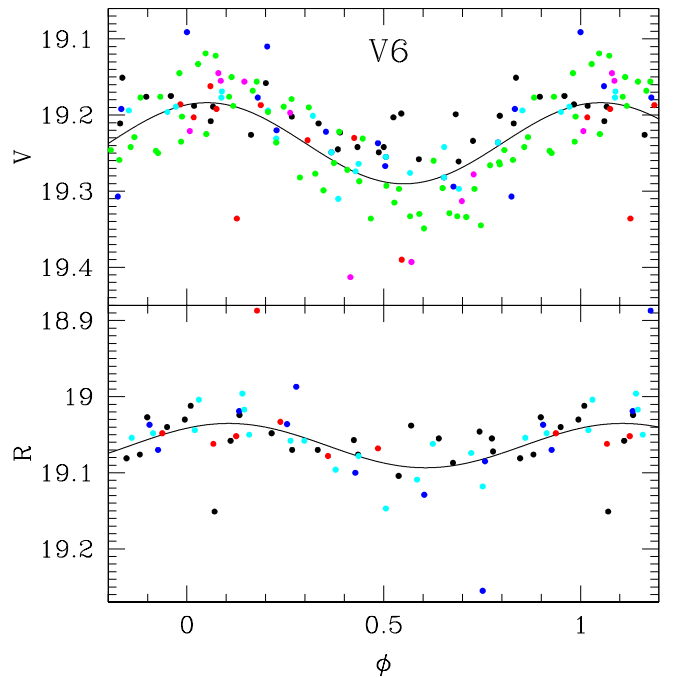}
\caption{Light curve of the newly discovered SX Phe star V6 in the $V$ filter (top) and $R$ filter (bottom) phased with the period 0.0565500~d. The
data point colours represent the different epochs listed in Table \ref{tab:colour_code}. The solid black curves represent the best fit sine
curves at the phasing period. The typical uncertainties in the $V$ and $R$ magnitudes are $\sim$0.02~mag.}
    \label{fig:V6}
\end{center}
\end{figure}

In figure \ref{fig:V6}, we present the phased light curve in the $V$ and $R$ filters. We overplot the best fit sine curve as a solid black line. As
expected, the $R$ filter light curve shows variations with the same period and phase as the $V$ filter light curve but with smaller amplitude.
There is a hint that the amplitude of V6 changed between different observing runs (compare the black points from 2004 with the green points from
2012). There are previous studies about SX Phe stars that show period change and also amplitude change. See for example figure 25
and section 4.2 in \cite{nemec95+03} and also section 5.2 in \cite{af10+02}.

SX Phe stars are well known as distance indicators through their P-L relation \citep[e.g.][]{jeon03+03}. By adopting the P-L relation for the
fundamental mode recently calculated by \cite{cohen12+01} for a sample of 77 double mode SX Phe stars in Galactic globular clusters, which is of the
form $M_{V}=-1.640(\pm0.110)-3.389(\pm0.090)\log(P_{f})$, we may calculate $M_V=$2.588$\pm$0.157~mag for V6 assuming that it is pulsating in the
fundamental mode. Given this, and assuming $E(B-V)=$0.0~mag, we obtain a true distance modulus $\mu_0=$16.644$\pm$0.157~mag, which translates to a 
distance of $\sim$21.3$\pm$1.5~kpc. Hence, if V6 is pulsating in the fundamental mode, it cannot be a cluster member.

However, if we assume that V6 is pulsating in the first overtone (1H), then we may ``fundamentalise'' the detected frequency by multiplying it by the
frequency ratio $f_1/f_2=$0.783 \citep[see][]{santolomazza01+05,jeon03+03, poretti05+15}. Using the \cite{cohen12+01} P-L
relation as before, we obtain $M_V=$2.228$\pm$0.149~mag, $\mu_0=$17.004$\pm$0.149~mag, and a distance of $\sim$25.2$\pm$1.8~kpc. Hence, if V6 is
pulsating in the first overtone, then it is most likely a cluster member.

Unfortunately, without detecting two frequencies in the light curve of V6, we cannot further speculate on the pulsation mode of this star.

{\bf V7}. Again, this variable was found above our detection threshold for the $\cal S_B$ statistic. In the CMD, it lies on the RGB at the edge of
the blue straggler region.  We analysed the $V$ light curve with {\tt Period04} and found a candidate frequency at 13.776775~cycles~d$^{-1}$ (or
a period of 0.072586~d). We did not find any further significant frequencies.

\begin{figure}[!t]
\begin{center}
\includegraphics[scale=0.51]{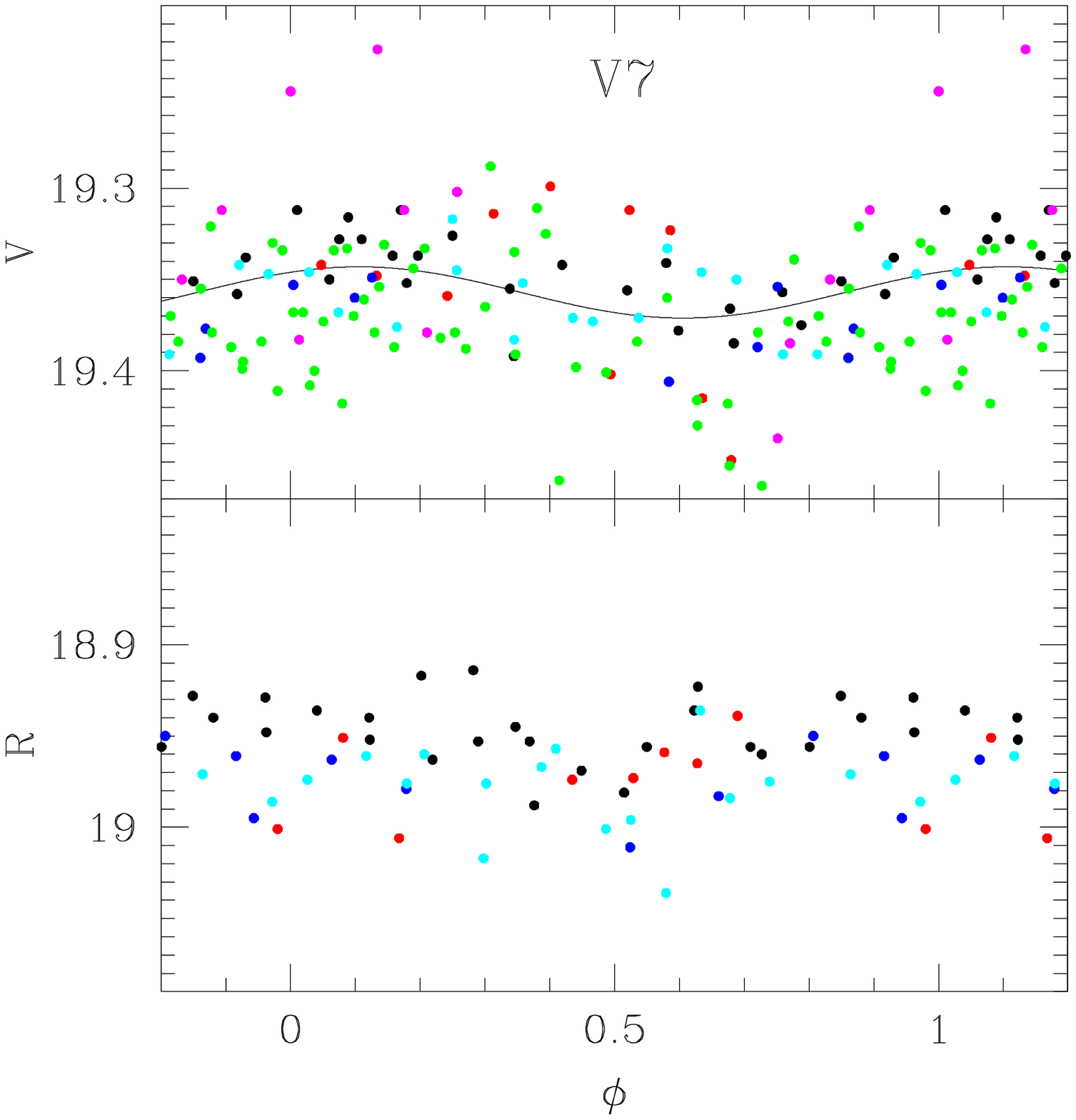}
\caption{Light curve of the newly discovered SX Phe star V7 in the $V$ filter (top) and $R$ filter (bottom) phased with the period 0.0725859~d. The
data point colours represent the different epochs listed in Table \ref{tab:colour_code}. The solid black curve represent the best fit sine
curve at the phasing period in the $V$ filter. The typical uncertainties in the $V$ and $R$ magnitudes are $\sim$0.02~mag.}
    \label{fig:V7}
\end{center}
\end{figure}

In figure \ref{fig:V7}, we present the phased light curve in the $V$ and $R$ filters along with the best fit sine curve in $V$ (solid black curve).
The variations at an amplitude of $\sim$0.05~mag are barely visible in the phased $V$ light curve. In order to quantify our classification of V7 as a
variable with the detected period, we calculate the improvement in chi-squared $\Delta \chi^{2}$ when fitting the sine curve compared to a constant
magnitude. Under the null hypothesis that the light curve is not variable, the $\Delta \chi^{2}$ statistic follows a chi-squared distribution with two
degrees of freedom. We set our threshold for rejection of the null hypothesis at 1\%, which is equivalent to $\Delta \chi^{2} \ga 9.21$. The $V$ light
curve of V7 has $\Delta \chi^{2} \approx 14.40$ which supports our conclusion that it is variable. We note that in the $R$ filter, the light curve is
not detected as showing variability by the $\Delta \chi^{2}$ test.

Using the \cite{cohen12+01} P-L relation for the detected frequency, we obtain $M_V=$2.221$\pm$0.152~mag, $\mu_0=$17.142$\pm$0.152~mag, and a distance
of $\sim$26.8$\pm$1.8~kpc, which is consistent with the distance to the cluster. Considering all the evidence we have discussed, we classify this
star as an SX Phe star that is most likely a cluster member pulsating in the fundamental mode.

\begin{figure}
\begin{center}
\includegraphics[scale=0.465]{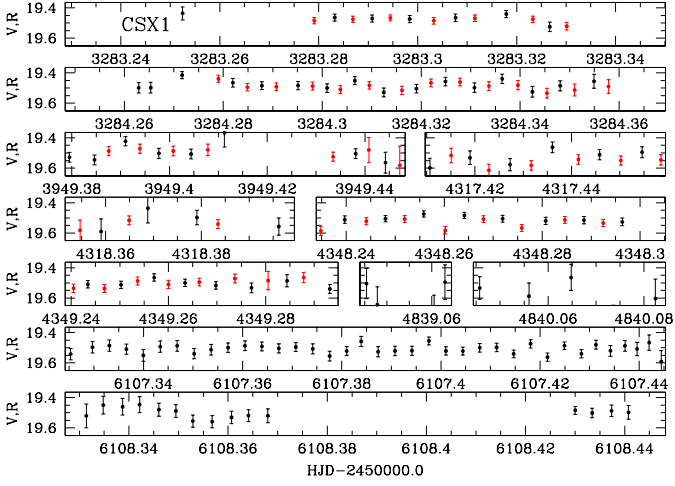}
\caption{Light curve of the candidate SX Phe star CSX1. Black points correspond to the $V$ magnitudes and red points correspond to
the $R$ magnitudes (which are shifted in mean magnitude to match the mean magnitude of the $V$ data). Some data points fall outside of the
plot magnitude range.}
\label{fig:CSXPhe}
\end{center}
\end{figure}

{\bf CSX1}. This star was detected above the threshold for the $\cal S_B$ statistic, and it lies in the blue straggler region in the CMD. We searched
for frequencies in the $V$ light curve using {\tt period04}, but we found no clear peaks. However, on our inspection of the light curve on individual
nights, we found clear cyclical variations on the time scale of $\sim$1~hour (see figure \ref{fig:CSXPhe}). Since we have been unable to detect a
pulsation frequency, we classify this star as a candidate SX Phe for which follow-up observations would be desirable.

Periods and amplitudes for the two new SX Phe and one candidate are given in Table \ref{tab:SXPHE}.  They are also labelled in the CMD of
Fig. \ref{fig:CMD} and their equatorial coordinates (J2000) are listed in the Table \ref{tab:astrom}. We note that the (V-R) colours of V6 and
V7, converted to (B-V) using the colour transformations of \cite{VandenBerg03+01}, are consistent (within 1$\sigma$) with the (B-V) colour-period
relation of \cite{McNamara11}.

\begin{table}
\caption{Detected pulsation frequencies for the new SX Phe variables discovered in NGC 7492. The numbers in parenthesis indicate the uncertainty on
the last decimal place.}
\centering
\begin{tabular}{lccccl}
\hline

ID   &    $A_0$\tablefootmark{a}  & Label  & Frequency   &  $A_V$\tablefootmark{b}   & mode \\
     &($V$ mag)  &        &(c/d$^{-1}$) &  (mag)   &      \\
\hline
 V6  & 19.235(4) & $f_1$  &17.683477(13)& 0.123(10)& $1H$?\tablefootmark{c} \\
 V7  & 19.363(4) & $f_1$  &13.776775(32)& 0.030(10) & $F$?\tablefootmark{c}\\
\hline
\end{tabular}
\tablefoot{\tablefoottext{a}{Mean V magnitude $A_0$}
\tablefoottext{b}{Full amplitude $A_V$ in the $V$ filter}
\tablefoottext{c}{If we assume that these stars are cluster members, then they are likely to be pulsating in the suggested mode.}}
\label{tab:SXPHE}
\end{table}

%
%
%

\section{Conclusions}
\label{sec:Concl}

Precise time series differential CCD $V, R, I$ photometry with a baseline of $\sim$8~years has been performed to detect brightness variations in
stars with $14.0<V<19.5$~mag in the field of NGC~7492. We found the following:

\begin{enumerate}

\item We identified one new long period variable (V5) and two SX Phe stars (V6 and V7). We present one candidate SX Phe (CSX1), which requires more
data of high precision to finally establish its nature. 

\item With the $\cal S_B$ variability statistic, it was possible to recover all previously known variables and also to find the new variables
presented in this work.

\item Our photometric precision at the magnitude of the horizontal branch combined with the consideration of the CMD means that we can be sure that
there are no undetected RR Lyrae stars that are cluster members in the field of view of our images.

\item For the RR0 V1, we improve the period estimate and perform a Fourier analysis to estimate a cluster metallicity of
[Fe/H]$_{\rm{ZW}}=$-1.68$\pm$0.10
 or [Fe/H]$_{\rm{UVES}}=$-1.64$\pm$0.13 and a distance of $\sim$24.3$\pm$0.5~kpc.

\item We found that the RR1 star V2 is undergoing a period change at a rate of $\beta\approx$47~d~Myr$^{-1}$. We also found tentative evidence for the
presence of the Blazhko effect in the light curve.

\item By assuming that the SX Phe stars are cluster members (which is consistent with their position in the CMD), we have used the SX Phe P-L
relation to speculate on the mode of oscillation of each star. We also obtain independent distance estimates to the cluster of $\sim$25.2$\pm$1.8
and 26.8$\pm$1.8~kpc. 

\item The cluster metallicity and distance estimates that we derive in this paper are all consistent with previous estimates in the literature.

\end{enumerate}

\begin{acknowledgements}
AAF acknowledges financial support from DGAPA-UNAM grant through project IN104612. We are thankful to the CONACyT (M\'exico) and the Department of
Science and Technology (India) for financial support under the Indo-Mexican collaborative project DST/INT/MEXICO/RP001/2001. We thank the staff at IAO
and at the remote control station at CREST, Hosakote for assistance during the observations. This work has made a large use of the SIMBAD and ADS
services, for which we are thankful.
\end{acknowledgements}

 \bibliography{psrrefs1}

\end{document}